\documentclass[prd,showpacs,nofootinbib,superscriptaddress,twocolumn]{revtex4}

\usepackage[latin1]{inputenc}
\usepackage{amsmath}
\usepackage{amsfonts}
\usepackage{amssymb}
\usepackage{pictex}

\def\be{\nopagebreak[3]\begin{equation}}
\def\ee{\end{equation}}
\def\ba{\nopagebreak[3]\begin{eqnarray}}
\def\ea{\end{eqnarray}}

\def\d{{\rm d}}

\def\a{\alpha}

\def\H{{\cal H}}

\def\R{\mathbb{R}}
\def\C{\mathbb{C}}
\def\Z{\mathbb{Z}}

\newcommand{\teta}{\rlap{\lower2ex\hbox{$\,\tilde{}$}}\eta{}}

\begin{document}
\preprint{\vbox{\baselineskip=12pt \rightline{IGPG-07/03-2}
\rightline{arXiv:0704.0007v2 [gr-qc]} }}
\title{Polymer Quantum Mechanics and its Continuum Limit}
\author{Alejandro Corichi}
\email{corichi@matmor.unam.mx} \affiliation{Instituto de
Matem\'aticas, Unidad Morelia, Universidad Nacional Aut\'onoma de
M\'exico, UNAM-Campus Morelia, A. Postal 61-3, Morelia,
Michoac\'an 58090, Mexico}

\affiliation{
Instituto de Ciencias Nucleares,
Universidad Nacional Aut\'onoma de M\'exico,\\
A. Postal 70-543, M\'exico D.F. 04510, Mexico}
\affiliation{Institute for Gravitational Physics and Geometry,
Physics Department, Pennsylvania State University, University Park
PA 16802, USA}
\author{Tatjana Vuka{\v s}inac}
\email{tatjana@shi.matmor.unam.mx} \affiliation{Facultad de
Ingenier\'\i a Civil, Universidad Michoacana de San Nicolas de
Hidalgo,\\
 Morelia, Michoac\'an 58000, Mexico}
\author{Jos\'e A. Zapata}\email{zapata@matmor.unam.mx}
\affiliation{Instituto de Matem\'aticas, Unidad Morelia,
Universidad Nacional Aut\'onoma de M\'exico, UNAM-Campus Morelia,
A. Postal 61-3, Morelia, Michoac\'an 58090, Mexico}

\begin{abstract}
A rather non-standard quantum representation of the canonical
commutation relations of quantum mechanics systems, known as the
{\it polymer representation} has gained some attention in recent
years, due to its possible relation with Planck scale physics. In
particular, this approach has been followed in a symmetric sector
of loop quantum gravity known as loop quantum cosmology. Here we
explore different aspects of the relation between the ordinary
Schr\"odinger theory and the polymer description. The paper has
two parts. In the first one, we derive the polymer quantum
mechanics starting from the ordinary Schr\"odinger theory and show
that the polymer description arises as an appropriate limit. In
the second part we consider the continuum limit of this theory,
namely, the reverse process in which one starts from the discrete
theory and tries to recover back the ordinary Schr\"odinger
quantum mechanics. We consider several examples of interest,
including the harmonic oscillator, the free particle and a simple
cosmological model.
\end{abstract}

\maketitle

\section{Introduction}

The so-called {\it polymer quantum mechanics}, a non-regular and
somewhat `exotic' representation of the canonical commutation
relations (CCR) \cite{non-reg}, has been used to explore both
mathematical and physical issues in background independent
theories such as quantum gravity \cite{AFW,freden}. A notable
example of this type of quantization, when applied to
minisuperspace models has given way to what is known as loop
quantum cosmology \cite{lqc,HW}. As in any toy model situation,
one hopes to learn about the subtle technical and conceptual
issues that are present in full quantum gravity by means of
simple, finite dimensional examples. This formalism is not an
exception in this regard. Apart from this motivation coming from
physics at the Planck scale, one can independently ask for the
relation between the standard continuous representations and their
polymer cousins at the level of mathematical physics. A deeper
understanding of this relation becomes important on its own.

The polymer quantization is made of several steps. The first one
is to build a representation of the Heisenberg-Weyl algebra on a
Kinematical Hilbert space that is ``background independent", and
that is sometimes referred to as the polymeric Hilbert space
$\H_{\rm poly}$. The second and most important part, the
implementation of dynamics, deals with the definition of a
Hamiltonian (or Hamiltonian constraint) on this space. In the
examples studied so far, the first part is fairly well understood,
yielding the kinematical Hilbert space $\H_{\rm poly}$ that is,
however, non-separable. For the second step, a natural
implementation of the dynamics has proved to be a bit more
difficult, given that a direct definition of the Hamiltonian
$\hat{H}$ of, say, a particle on a potential on the space $\H_{\rm
poly}$ is not possible since one of the main features of this
representation is that the operators $\hat{q}$ and $\hat{p}$
cannot be both simultaneously defined (nor their analogues in
theories involving more elaborate variables). Thus, any operator
that involves (powers of) the not defined variable has to be
regulated by a well defined operator which normally involves
introducing some extra structure on the configuration (or
momentum) space, namely a lattice. However, this new structure
that plays the role of a regulator can not be removed when working
in $\H_{\rm poly}$ and
one is left with the ambiguity that is present in any
regularization. The freedom in choosing it can be sometimes
associated with a length scale (the lattice spacing). For ordinary
quantum systems such as a simple harmonic oscillator, that has
been studied in detail from the polymer viewpoint, it has been
argued that if this length scale is taken to be `sufficiently
small', one can arbitrarily approximate standard Schr\"odinger
quantum mechanics \cite{AFW,freden}. In the case of loop quantum
cosmology, the minimum area gap $A_0$ of the full quantum gravity
theory imposes such a scale, that is then taken to be fundamental
\cite{lqc}.

A natural question is to ask what happens when we change this
scale and go to even smaller `distances', that is, when we refine
the lattice on which the dynamics of the theory is defined. Can we
define consistency conditions between these scales? Or even
better, can we take the limit and find thus a continuum limit? As
it has been shown recently in detail, the answer to both questions
is in the affirmative \cite{CVZ}. There, an appropriate notion of
scale was defined in such a way that one could define refinements
of the theory and pose in a precise fashion the question of the
continuum limit of the theory. These results could also be seen as
handing a procedure to remove the regulator when working on the
appropriate space. The purpose of {\it this} paper is to further
explore different aspects of the relation between the continuum
and the polymer representation. In particular in the first part we
put forward a novel way of deriving the polymer representation
from the ordinary  Schr\"odinger representation as an appropriate
limit. In Sec.~\ref{sec:2} we derive two versions of the polymer
representation as different limits of the Schr\"odinger theory. In
Sec.~\ref{sec:4} we show that these two versions can be seen as
different polarizations of the `abstract' polymer representation.
These results, to the best of our knowledge, are new and have not
been reported elsewhere. In Sec.~\ref{sec:5} we pose the problem
of implementing the dynamics on the polymer representation. In
Sec.~\ref{sec:6} we motivate further the question of the continuum
limit (i.e. the proper removal of the regulator) and recall the
basic constructions of \cite{CVZ}. Several examples are considered
in Sec.~\ref{sec:7}. In particular a simple harmonic oscillator,
the polymer free particle and a simple quantum cosmology model are
considered. The free particle and the cosmological model represent
a generalization of the results obtained in \cite{CVZ} where only
systems with a discrete and non-degenerate spectrum where
considered. We end the paper with a discussion in
Sec.~\ref{sec:8}. In order to make the paper self-contained, we
will keep the level of rigor in the presentation to that found in
the standard theoretical physics literature.

\section{Quantization and Polymer Representation}
\label{sec:2}

In this section we derive the so called polymer representation of
quantum mechanics starting from a specific reformulation of the
ordinary Schr\"odinger representation. Our starting point will be
the simplest of all possible phase spaces, namely $\Gamma=\R^2$
corresponding to a particle living on the real line $\R$. Let us
choose coordinates $(q,p)$ thereon.
As a first step we shall
consider the quantization of this system that leads to the
standard quantum theory in the Schr\"odinger description. A
convenient route is to introduce the necessary structure to define
the Fock representation of such system. From this perspective, the
passage to the polymeric case becomes clearest. Roughly speaking
by a quantization one means a passage from the classical algebraic
bracket, the Poisson bracket,
\be
 \{q,\,p\}=1
\ee
to a quantum bracket given by the commutator of the corresponding
operators, \be [\,\hat{q},\,\hat{p}]=i\hbar\,\hat{{\mathbf 1}} \ee
These relations, known as the canonical commutation relation
(CCR) become the most common corner stone of the (kinematics of the) quantum
theory; they should be satisfied by
the quantum system, when represented on a Hilbert space ${\H}$.

There are alternative points of departure for quantum kinematics.
Here we consider the algebra generated by the exponentiated versions
of $\hat{q}$ and $\hat{p}$ that are denoted by,
$$
U(\alpha)=e^{i(\alpha \,\hat{q})/\hbar} \qquad;\qquad
V(\beta)=e^{i(\beta\, \hat{p})/\hbar}
$$
where $\alpha$ and $\beta$ have dimensions of momentum and length,
respectively. The CCR now become \be U(\alpha)\cdot
V(\beta)=e^{(-i\alpha\,\beta)/\hbar}V(\beta)\cdot U(\alpha)
\label{W-P} \ee and the rest of the product is
$$
U(\alpha_1)\cdot U(\alpha_2)=U(\alpha_1+\alpha_2)\quad ;\quad
V(\beta_1)\cdot V(\beta_2)=V(\beta_1+\beta_2)
$$
The Weyl algebra ${\cal W}$ is generated by taking finite linear combinations
of the
generators $U(\alpha_i)$ and $V(\beta_i)$ where the product (\ref{W-P}) is
extended by linearity,
$$
\sum_{i} (A_i\,U(\alpha_i)+B_i\,V(\beta_i))
$$
From this perspective,  quantization means finding an unitary
representation of the Weyl algebra ${\cal W}$ on a Hilbert space
$\H'$ (that could be different from the ordinary Schr\"odinger
representation). At first it might look weird to attempt this
approach given that we {\em know} how to quantize such a simple
system; what do we need such a complicated object as ${\cal W}$
for? It is infinite dimensional, whereas the set ${\cal
S}=\{\hat{\mathbf 1},\hat{q},\hat{p}\}$, the starting point of the
ordinary Dirac quantization, is rather simple. It is in the
quantization of field systems that the advantages of the Weyl
approach can be fully appreciated, but it is also useful for
introducing the polymer quantization and comparing it to the
standard quantization. This is the strategy that we follow.

A question that one can ask is whether there is any freedom in quantizing the
system to obtain the ordinary Schr\"odinger representation. On a first sight it
might seem that there is none given the Stone-Von Neumann uniqueness theorem.
Let us review what would be the argument for the
standard construction. Let us ask that the representation we want to build up
is of the Schr\"odinger type, namely, where states are wave functions of
configuration space $\psi(q)$. There are two ingredients to the construction
of the representation, namely the specification of how the basic operators
$(\hat{q},\hat{p})$ will act, and the nature of the space of functions that
$\psi$ belongs to, that is normally fixed by the choice of inner product on
$\H$,
or measure $\mu$ on $\R$. The standard choice is to select the Hilbert space to
be,
$$
\H=L^2(\R,\d q)
$$
the space of square-integrable functions with respect to the Lebesgue measure
$\d q$ (invariant under constant translations) on $\R$. The operators are then
represented as,
\be
\hat{q}\cdot\psi(q)=(q\,\psi)(q)\quad {\rm and}\quad \hat{p}\cdot\psi(q)=
-i\,\hbar\,\frac{\partial}{\partial q}\,\psi(q)\label{simpl-r}
\ee
Is it possible to find other representations?
In order to appreciate this freedom we go to the
Weyl algebra and build the quantum theory thereon. The representation of the
Weyl algebra that
can be called of the `Fock type' involves the definition of an extra structure
on the phase space $\Gamma$: a complex structure $J$. That is, a linear mapping
 from $\Gamma$ to itself such that $J^2=-1$. In 2 dimensions, all the freedom
in the choice of $J$ is contained in the choice of a  parameter
$d$ with dimensions of length. It is also convenient to define: $k=p/\hbar$
that has dimensions of $1/L$.
We have then,
$$
J_d:(q,k)\mapsto (-d^2\,k,q/d^2)
$$
This object together with the symplectic structure: $\Omega((q,p);(q',p'))=
q\,p'-p\,q'$ define an inner product on $\Gamma$ by the formula
$g_d(\cdot\,;\,\cdot)=\Omega(\cdot\,;J_d\,\cdot)$ such that:
$$
g_d((q,p);(q',p'))=\frac{1}{d^2}\,q\,q'+\frac{d^2}{\hbar^2}\,p\,p'
$$
which is dimension-less and positive definite. Note that with this quantities
one can define complex coordinates $(\zeta,\bar{\zeta})$ as usual:
$$
\zeta=\frac{1}{d}\,q + i\frac{d}{\hbar}\,p\quad ;\quad
\bar{\zeta}=\frac{1}{d}\,q - i\frac{d}{\hbar}\,p
$$
from which one can build the standard Fock representation. Thus,
one can alternatively view the introduction of the length
parameter $d$ as the quantity needed to define (dimensionless)
complex coordinates on the phase space. But what is the relevance
of this object ($J$ or $d$)? The definition of complex
coordinates is useful for the construction of the Fock space since
from them one can define, in a natural way, creation and
annihilation operators. But for the Schr\"odinger representation
we are interested here, it is a bit more subtle. The subtlety is
that within this approach one uses the algebraic properties of
${\cal W}$ to construct the Hilbert space via what is known as the
Gel'fand-Naimark-Segal (GNS) construction. This implies that the
measure in the Schr\"odinger representation becomes non trivial
and thus the momentum operator acquires an extra term in order to
render the operator self-adjoint. The representation of the Weyl
algebra is then, when acting on functions $\phi(q)$ \cite{AC:JC}:
$$
\hat{U}(\alpha)\cdot\phi(q):=(e^{i\alpha\,q/\hbar}\,\phi)(q)
$$
and,
$$
\hat{V}(\beta)\cdot\phi(q):=e^{\frac{\beta}{d^2}(q-\beta/2)}\,\phi(q-\beta)
$$
The Hilbert space structure is introduced by the definition of an
algebraic state (a positive linear functional) $\omega_d:{\cal
W}\rightarrow \C$, that must coincide with the expectation
value in the Hilbert space taken on a special state
refered to as the vacuum: $\omega_d(a)=\langle\hat{a}\rangle_{\rm vac}$,
for all $a\in {\cal W}$. In our case this  specification of $J$
induces such a unique state $\omega_d$ that yields,
\be
\langle \hat{U}(\alpha)\rangle_{\rm vac}=e^{-\frac{1}{4}\frac{d^2\,\alpha^2}
{\hbar^2}}\label{a-state-1}
\ee and
\be
\langle \hat{V}(\beta)\rangle_{\rm vac}=e^{-\frac{1}{4}\frac{\beta^2}{d^2}}
\label{a-state-2}
\ee
Note that the exponents in the vacuum expectation values
correspond to the metric constructed out of $J$:
$\frac{d^2\,\alpha^2}{\hbar^2}=g_d((0,\alpha);(0,\alpha))$ and
$\frac{\beta^2}{d^2}=g_d((\beta,0);(\beta,0))$. Wave functions
belong to the space $L^2(\R,\d\mu_d)$, where the measure that
dictates the inner product in this representation is given by,
$$
\d\mu_d=\frac{1}{d\sqrt{\pi}}\,e^{-\frac{q^2}{d^2}}\,\d q
$$
In this representation, the vacuum is given by the identity function
$\phi_0(q)=1$ that is, just as any plane wave, normalized.
Note that for each value of $d>0$, the representation is well defined and
continuous in $\alpha$ and $\beta$. Note also that there is an equivalence
between the $q$-representation
defined by $d$ and the $k$-representation defined by $1/d$.

How can we recover then the standard representation in which the measure is
given by the Lebesgue measure and the operators are represented as in
(\ref{simpl-r})?
It is easy to see that there is an isometric isomorphism $K$ that maps the
$d$-representation in $\H_d$ to the standard Schr\"odinger representation in
$\H_{\rm schr}$ by:
$$
\psi(q)=K\cdot \phi(q)=\frac{e^{-\frac{q^2}{2\,d^2}}}{d^{1/2}\pi^{1/4}}
\,\phi(q)\,\in\, \H_{\rm schr}=L^2(\R,\d q)
$$
Thus we see that all $d$-representations are unitarily equivalent.
This was to be expected in view of the Stone-Von Neumann
uniqueness result. Note also that the vacuum now becomes
$$
\psi_0(q)=\frac{1}{d^{1/2}\pi^{1/4}}\,e^{-\frac{q^2}{2\,d^2}}\, ,
$$
so even when there is no information about the parameter $d$ in
the representation itself, it is contained in the vacuum state.
This procedure for constructing the GNS-Schr\"odinger
representation for quantum mechanics has also been generalized to
scalar fields on arbitrary curved space in \cite{CCQ}. Note,
however that so far the treatment has all been kinematical,
without any knowledge of a Hamiltonian. For the Simple Harmonic
Oscillator of mass $m$ and frequency $\omega$, there is a natural
choice compatible with the dynamics given by
$d=\sqrt{\frac{\hbar}{m\,\omega}}$, in which some calculations simplify
(for instance for coherent states), but in principle one can use any value of
$d$.

Our study will be simplified by focusing on
the fundamental entities in
the Hilbert Space $\H_d$ , namely those states generated by acting
with $\hat{U}(\alpha)$ on the vacuum $\phi_0(q)=1$. Let us denote those states
by,
$$
\phi_\alpha(q)=\hat{U}(\alpha)\cdot\phi_0(q)=e^{i\frac{1}{\hbar}\alpha\,q}
$$
The inner product between two such states is given by
\be
\langle\phi_\alpha,\phi_\lambda\rangle_d=\int\d\mu_d\;
e^{-\frac{i\alpha q}{\hbar}}\,
e^{\frac{i\lambda q}{\hbar}}=e^{-\frac{(\lambda-\alpha)^2\,d^2}{4\,\hbar^2}}
\label{d-IP}
\ee
Note incidentally that, contrary to some common belief, the `plane waves' in
this
GNS Hilbert space are indeed normalizable.

Let us now consider the polymer representation. For that, it is important to
note that there are two possible limiting cases for the parameter $d$: i) The
limit $1/d\mapsto 0$ and ii) The case $d\mapsto 0$. In both cases, we have
expressions that become ill defined in the representation or measure, so one
needs to be careful.


\subsection{The $1/d \mapsto 0$ case.}

\noindent
The first observation is that from the expressions (\ref{a-state-1}) and
(\ref{a-state-2}) for the algebraic state $\omega_d$, we see that the limiting
cases are indeed well defined. In our case we get,
$\omega_{\rm A}:=\lim_{1/d\to 0}\omega_d$ such that,
\be
\omega_{\rm A}(\hat{U}(\alpha))=\delta_{\alpha,0}\qquad{\rm and}\qquad
\omega_{\rm A}(\hat{V}(\beta))=1
\ee
From this, we can indeed construct the representation by means of the GNS construction. In order to do that and to
show how this is obtained we shall consider several expressions. One has to be
careful though, since the limit has to be taken with care.
Let us consider the measure on the representation that behaves as:
$$
\d\mu_d=\frac{1}{d\sqrt{\pi}}\,e^{-\frac{q^2}{d^2}}\,\d q\mapsto
\frac{1}{d\sqrt{\pi}}\;\d q
$$
so the measures tends to an homogeneous measure but whose `normalization
constant' goes to zero, so the limit becomes somewhat subtle. We shall return
to this point later.

Let us now see what happens to the inner product between the fundamental
entities in the Hilbert Space $\H_d$ given by (\ref{d-IP}).
It is immediate to see that in the $1/d\mapsto 0$ limit the inner
product becomes, \be
\langle\phi_\alpha,\phi_\lambda\rangle_d\mapsto
\delta_{\alpha,\lambda}
\ee
with $\delta_{\alpha,\lambda}$ being Kronecker's delta. We see
then that the plane waves $\phi_\alpha(q)$ become an orthonormal
basis for the new Hilbert space. Therefore, there is a delicate
interplay between the two terms that contribute to the measure in
order to maintain the normalizability of these functions; we need
the measure to become damped (by $1/d$) in order to avoid that the
plane waves acquire an infinite norm (as happens with the standard
Lebesgue measure), but on the other hand the measure, that for any
finite value of $d$ is a Gaussian, becomes more and more spread.

It is important to note that, in this limit, the operators $\hat{U}(\alpha)$
become discontinuous with respect to $\alpha$, given that for any given
$\alpha_1$ and $\alpha_2$ (different), its action on a given basis vector
$\psi_\lambda(q)$ yields orthogonal vectors. Since the continuity of these
operators is one of the hypotesis of the Stone-Von Neumann theorem, the
uniqueness result does not apply here. The representation is inequivalent
to the standard one.

Let us now analyze the other operator, namely the action of the operator
$\hat{V}(\beta)$ on the basis $\phi_\alpha(q)$:
\[
\hat{V}(\beta)\cdot\phi_\alpha(q)=e^{-\frac{\beta^2}{2d^2}-i
\frac{\alpha\beta}{\hbar}}\;
e^{(\beta/d^2+i\alpha/\hbar)q}
\]
which in the limit $1/d\mapsto 0$ goes to,
\[
\hat{V}(\beta)\cdot\phi_\alpha(q)\mapsto e^{i\frac{\alpha\beta}{\hbar}}\,
\phi_\alpha(q)
\]
that is continuous on $\beta$. Thus, in the limit, the operator
$\hat{p}=-i\hbar\partial_q$ is well defined. Also, note that in this limit the
operator $\hat{p}$ has $\phi_\alpha(q)$ as its eigenstate with eigenvalue
given by
$\alpha$:
\[
\hat{p}\cdot\phi_\alpha(q)\mapsto\alpha\,\phi_\alpha(q)
\]
To summarize, the resulting theory obtained by taking the limit $1/d\mapsto 0$
of the ordinary Schr\"odinger description, that we shall call the `polymer
representation of type A', has the following features:
the operators $U(\alpha)$ are well defined but not continuous in $\alpha$,
so there is no
generator (no operator associated to $q$). The basis vectors $\phi_\alpha$
are orthonormal (for $\alpha$ taking values on a continuous set) and are
eigenvectors of the operator
$\hat{p}$ that {\it is} well defined. The resulting Hilbert space $\H_A$ will
be the (A-version of the) polymer representation.
Let us now consider the other case, namely, the limit when $d\mapsto 0$.

\subsection{The $d \mapsto 0$ case}

\noindent Let us now explore the other limiting case of the
Schr\"odinger/Fock representations labelled by the parameter $d$.
Just as in the previous case, the limiting algebraic state
becomes, $\omega_{\rm B}:=\lim_{d\to 0}\omega_d$ such that, \be
\omega_{\rm B}(\hat{U}(\alpha))=1\qquad{\rm and}\qquad \omega_{\rm
B}(\hat{V}(\beta))=\delta_{\beta,0}
\ee
From this positive linear function, one can indeed construct the
representation using the GNS construction.

First let us note that the measure, even when the limit has to be
taken with due care, behaves as:
$$
\d\mu_d=\frac{1}{d\sqrt{\pi}}\,e^{-\frac{q^2}{d^2}}\,\d q\mapsto \delta(q)\;\d
q
$$
That is, as Dirac's delta distribution.
It is immediate to see that, in the $d \mapsto 0$ limit, the inner product
between the fundamental states $\phi_\alpha (q)$ becomes,
\be
\langle\phi_\alpha,\phi_\lambda\rangle_d\mapsto 1
\ee
This in fact means that the vector $\xi=\phi_\alpha-
\phi_\lambda$ belongs to the Kernel of the limiting inner product, so one has
to mod out by these (and all) zero norm states in order to get the Hilbert
space.

Let us now analyze the other operator, namely the action of the operator
$\hat{V}(\beta)$ on the vacuum $\phi_0(q)=1$, which for arbitrary $d$ has
the form,
\[
\tilde{\phi}_\beta:=\hat{V}(\beta)\cdot\phi_0(q)=e^{\frac{\beta}{d^2}(q-\beta/2)}
\]
The inner product between two such states is given by
$$\langle \tilde{\phi}_\alpha , \tilde{\phi}_\beta\rangle_d\ =
e^{-\frac{1}{4d^2}(\alpha -\beta )^2}$$
In the limit $d\to 0$, $\langle \tilde{\phi}_\alpha , \tilde{\phi}_\beta
\rangle_d\to\delta_{\alpha ,\beta}$. We can see then that it is these functions
that become the orthonormal, `discrete basis' in the theory. However,
the function $\tilde{\phi}_\beta (q)$ in this limit becomes ill defined. For
example, for $\beta >0$, it grows
unboundedly for $q>\beta/2$, is equal to one if $q=\beta/2$ and
zero otherwise. In order to overcome these difficulties and make more
transparent
the resulting theory, we shall
consider the other form of the representation in which the measure
is incorporated into the states (and the resulting Hilbert space
is $L^2(\R,\d q)$). Thus the new state
\ba \psi_\beta(q) &:=& K\cdot
(\hat{V}(\beta)\cdot\phi_0(q))= \nonumber\\
&=& \frac{1}{(d\sqrt{\pi})^\frac{1}{2}}\,
e^{-\frac{1}{2d^2}(q-\beta)^2}
\ea
We can now take the limit and what we get is
\[
\lim_{d\mapsto 0} \psi_\beta(q):=\delta^{1/2}(q,\beta)
\]
where by $\delta^{1/2}(q,\beta)$ we mean something like `the
square root of the Dirac distribution'. What we really mean is an
object that satisfies the following property:
\[
\delta^{1/2}(q,\beta)\cdot\delta^{1/2}(q,\alpha)=\delta(q,\beta)\,\delta_{\beta,\alpha}
\]
That is, if $\alpha=\beta$ then it is just the ordinary delta,
otherwise it is zero. In a sense these object can be regarded as
half-densities that can not be integrated by themselves, but whose
product can. We conclude then that the inner product is, \be
\langle \psi_\beta,\psi_\alpha\rangle=\int_\R\d
q\;\overline{\psi_\beta}(q)\,\psi_\alpha(q) =\int_\R\d q
\;\delta(q,\alpha)\,\delta_{\beta,\alpha}=\delta_{\beta,\alpha}\label{IP-B}
\ee which is just what we expected. Note that in this
representation, the vacuum state becomes
$\psi_0(q):=\delta^{1/2}(q,0)$, namely, the half-delta with
support in the origin. It is important to note that we are
arriving in a natural way to states as half-densities, whose
squares can be integrated without the need of a nontrivial measure
on the configuration space. Diffeomorphism invariance arises then
in a natural but subtle manner.

Note that as the end result we recover the Kronecker delta inner
product for the new fundamental states:
\[ \chi_\beta(q):=\delta^{1/2}(q,\beta).\]
Thus, in this new B-polymer representation, the Hilbert space
$\H_{\rm B}$ is the completion with respect to the inner product
(\ref{IP-B}) of the states generated by taking (finite) linear
combinations of basis elements of the form $\chi_\beta$: \be
\Psi(q)=\sum_i\,b_i\,\chi_{\beta_i}(q)
\ee
Let us now introduce an equivalent description of this Hilbert
space. Instead of having the basis elements be half-deltas as
elements of the Hilbert space where the inner product is given by
the ordinary Lebesgue measure $\d q$, we redefine both the basis
and the measure. We could consider, instead of a half-delta with
support $\beta$, a Kronecker delta or characteristic function with
support on $\beta$: $$ \chi'_\beta(q):=\delta_{q,\beta}$$ These
functions have a similar behavior with respect to the product as
the half-deltas, namely: $\chi'_\beta(q)\cdot
\chi'_\alpha(q)=\delta_{\beta,\alpha}$. The main difference is
that neither $\chi'$ nor their squares are integrable with respect
to the Lebesgue measure (having zero norm). In order to fix that
problem we have to change the measure so that we recover the basic
inner product (\ref{IP-B}) with our new basis. The needed measure
turns out to be the discrete counting measure on $\R$. Thus any
state in the `half density basis' can be written (using the same
expression) in terms of the `Kronecker basis'. For more details
and further motivation see the next section.

Note that in this B-polymer representation, both $\hat{U}$ and $\hat{V}$ have
their roles interchanged with that of the A-polymer representation: while
$U(\alpha)$ is discontinuous and thus $\hat{q}$ is not defined in the
A-representation, we have that it is $V(\beta)$ in the B-representation
that has this property. In this case, it is  the operator $\hat{p}$ that
can not be defined.
We see then that given a physical system for which the configuration space
has a well defined physical meaning, within the possible representation in
which wave-functions are functions of the configuration variable $q$, the A
and B polymer representations are radically different and inequivalent.

Having said this, it is also true that the A and B representations
are equivalent in a different sense, by means of the duality
between $q$ and $p$
representations and the $d\leftrightarrow 1/d$ duality: The
A-polymer representation in the ``$q$-representation" is
equivalent to the B-polymer representation in the
``$p$-representation", and conversely. When studying a problem, it
is important to decide from the beginning which polymer
representation (if any) one should be using (for instance in the
$q$-polarization). This has as a consequence an implication on
which variable is naturally ``quantized" (even if continuous): $p$
for A and $q$ for B. There could be for instance a physical
criteria for this choice. For example a fundamental symmetry could
suggest that one representation is more natural than another one.
This indeed has been recently noted by Chiou in \cite{chiou},
where the Galileo group is investigated and where it is shown that
the B representation is better behaved.

In the other polarization, namely for wavefunctions of $p$, the
picture gets reversed: $q$ is discrete for the A-representation,
while $p$ is for the B-case. Let us end this section by noting
that the procedure of obtaining the polymer quantization by means
of an appropriate limit of Fock-Schr\"odinger representations
might prove useful in more general settings in field theory or
quantum gravity.

\section{Polymer Quantum Mechanics: Kinematics}
\label{sec:4}

In previous sections we have derived what we have called the A and B polymer
representations (in the $q$-polarization) as limiting cases of ordinary Fock
representations.
In this section, we shall describe, without any reference to the Schr\"odinger
representation, the `abstract' polymer representation and then make contact
with its two possible realizations, closely related to the A and B cases
studied before.
What we will see is that one of them (the A case) will correspond to the
$p$-polarization while the other one corresponds to the $q-$representation,
when
a choice is made about the physical significance of the variables.

We can start by defining abstract kets $|\mu\rangle$  labelled by
a real number $\mu$. These shall belong to the Hilbert space
$\H_{\rm poly}$. From these states, we define a generic `cylinder
states' that correspond to a choice of a finite collection of
numbers $\mu_i\in \R$ with $i=1,2,\ldots, N$. Associated to this
choice, there are $N$ vectors $|\mu_i\rangle$, so we can take a
linear combination of them \be |\psi\rangle = \sum_{i=1}^N
a_i\,|\mu_i\rangle \label{lin-comb} \ee The polymer inner product
between the fundamental kets is given by, \be \langle
\nu|\mu\rangle=\delta_{\nu,\mu} \label{poly-IP2} \ee That is, the
kets are orthogonal to each other (when $\nu\neq\mu$) and they are
normalized ($\langle\mu|\mu\rangle=1$). Immediately, this implies
that, given any two vectors $|\phi\rangle =\sum_{j=1}^M
b_j|\nu_j\rangle$ and $|\psi\rangle=\sum_{i=1}^N
a_i\,|\mu_i\rangle$, the inner product between them is given by,
\[
\langle \phi|\psi\rangle=\sum_i\sum_j \,\bar{b}_j\,a_i\,\langle
\nu_j|\mu_i\rangle=
\sum_k \bar{b}_k\,a_{k}
\]
where the sum is over $k$ that labels the intersection points between the
set of labels $\{\nu_j\}$ and $\{\mu_i\}$. The Hilbert space $\H_{\rm poly}$
is the Cauchy completion of finite linear combination of the form
(\ref{lin-comb}) with respect to the inner product (\ref{poly-IP2}).
$\H_{\rm poly}$ is non-separable. There are two basic operators on this
Hilbert space: the `label operator' $\hat{\varepsilon}$:
\[
\hat{\varepsilon}\,|\mu\rangle := \mu\,|\mu\rangle
\]
and the displacement operator $\hat{\bf s}\,(\lambda)$,
$$
\hat{\bf s}\,(\lambda)\,|\mu\rangle:=|\mu+\lambda\rangle
$$
The operator  $\hat{\varepsilon}$ is symmetric and the operator(s)
$\hat{\bf s}(\lambda)$ defines a one-parameter family of unitary
operators on $\H_{\rm poly}$, where its adjoint is given by
$\hat{\bf s}^\dagger\,(\lambda)=\hat{\bf s}\,(-\lambda)$. This
action is however, discontinuous with respect to $\lambda$ given
that $|\mu\rangle$ and $|\mu+\lambda\rangle$ are always
orthogonal, no matter how small is $\lambda$. Thus, there is no
(Hermitian) operator that could generate $\hat{\bf s}\,(\lambda)$
by exponentiation.

So far we have given the abstract characterization of the Hilbert space,
but one would like to make contact with concrete realizations as wave
functions, or by identifying the abstract operators $\hat{\varepsilon}$ and
$\hat{\bf s}$ with physical operators.

Suppose we have a system with a configuration space with
coordinate given by $q$, and $p$ denotes its canonical conjugate
momenta. Suppose also that for physical reasons we decide that the
configuration coordinate $q$ will have some ``discrete character"
(for instance, if it is to be identified with position, one could
say that there is an underlying discreteness in position at a
small scale). How can we implement such requirements by means of
the polymer representation? There are two possibilities, depending
on the choice of `polarizations' for the wavefunctions, namely
whether they will be functions of configuration $q$ or momenta
$p$. Let us the divide the discussion into two parts.

\subsection{Momentum polarization}

\noindent
In this polarization, states will be denoted by,
$$
\psi(p)=\langle p|\psi\rangle
$$
where
$$
\psi_\mu(p)=\langle p|\mu\rangle=e^{i\frac{\mu p}{\hbar}}
$$
How are then the operators $\hat{\varepsilon}$ and $\hat{\bf s}$ represented?
Note that if we associate the multiplicative operator
$$
\hat{V}(\lambda)\cdot \psi_\mu(p)= e^{i\frac{\lambda\,p}{\hbar}}\,
e^{i\frac{\mu\,p}{\hbar}}=e^{i\,\frac{(\mu+\lambda)}{\hbar}\,p}=
\psi_{(\mu+\lambda)}(p)
$$
we see then that the operator $\hat{V}(\lambda)$ corresponds
precisely to the shift operator $\hat{\bf s}\,(\lambda)$. Thus we
can also conclude that the operator $\hat{p}$ does not exist. It
is now easy to identify the operator $\hat{q}$ with:
$$
\hat{q}\cdot\psi_\mu(p)=-i\hbar\,\frac{\partial}{\partial
p}\,\psi_\mu(p)=\mu\,e^{i\, \frac{\mu\,p}{\hbar}}=\mu\,\psi_\mu(p)
$$
namely, with the abstract operator $\hat{\varepsilon}$. The reason
we say that $\hat{q}$ is discrete is because this operator has as
its eigenvalue the label $\mu$ of the elementary state
$\psi_\mu(p)$, and this label, even when it can take value in a
continuum of possible values, is to be understood as a {\em
discrete} set, given that the states are orthonormal for all
values of $\mu$. Given that states are now functions of $p$, the
inner product (\ref{poly-IP2}) should be defined by a measure
$\mu$ on the space on which the wave-functions are defined. In
order to know what these two objects are, namely, the quantum
``configuration" space ${\cal C}$ and the measure
thereon\footnote{here we use the standard terminology of
`configuration space' to denote the domain of the wave function
even when, in this case, it corresponds to the physical momenta
$p$.}, we have to make use of the tools available to us from the
theory of $C^*$-algebras. If we consider the operators
$\hat{V}(\lambda)$, together with their natural product and
$*$-relation given by $\hat{V}^*(\lambda)=\hat{V}(-\lambda)$, they
have the structure of an Abelian $C^*$-algebra (with unit) ${\cal
A}$. We know from the representation theory of such objects that
${\cal A}$ is isomorphic to the space of continuous functions
$C^0(\Delta)$ on a compact space $\Delta$, the {\it spectrum} of
${\cal A}$. Any representation of ${\cal A}$ on a Hilbert space as
multiplication operator will be on spaces of the form
$L^2(\Delta,\d\mu)$. That is, our quantum configuration space is
the spectrum of the algebra, which in our case corresponds to the
{\it Bohr compactification} $\R_{\rm b}$ of the real line
\cite{Bohr}. This space is a compact group and there is a natural
probability measure defined on it, the Haar measure $\mu_{\rm H}$.
Thus, our Hilbert space $\H_{\rm poly}$ will be isomorphic to the
space, \be \H_{{\rm poly},p}=L^2(\R_{\rm b},\d\mu_{\rm H}) \ee In
terms of `quasi periodic functions' generated by $\psi_\mu(p)$,
the inner product takes the form
\ba
\langle \psi_\mu|\psi_\lambda\rangle &:=& \int_{\R_{\rm b}}\d\mu_{\rm H}\,
\overline{\psi_\mu}(p)\,\psi_\lambda(p):=\nonumber \\
&=& \lim_{L\mapsto
\infty}\frac{1}{2L} \int^L_{-L}\d p\,
\overline{\psi_\mu}(p)\,\psi_\lambda(p)=\delta_{\mu,\lambda}
\ea
note  that in the $p$-polarization, this characterization
corresponds to the `A-version' of the polymer representation of
Sec.~\ref{sec:2} (where $p$ and $q$ are interchanged).

\subsection{$q$-polarization}

\noindent
Let us now consider the other polarization in which wave functions will
depend on the configuration coordinate $q$:
$$
\psi(q)=\langle q|\psi\rangle
$$
The basic functions, that now will be called
$\tilde{\psi}_\mu(q)$, should be, in a sense, the dual of the
functions $\psi_\mu(p)$ of the previous subsection. We can try to
define them via a `Fourier transform':
\[
\tilde{\psi}_\mu(q) := \langle q|\mu\rangle=\langle q|
\int_{\R_{\rm b}} \d\mu_{\rm H} |p\rangle\langle
p|\mu\rangle \]
which is given by
\ba
\tilde{\psi}_\mu(q) &:=& \int_{\R_{\rm b}}\d\mu_{\rm H} \langle q|p\rangle
\psi_\mu(p)=\nonumber \\
&=& \int_{\R_{\rm b}}\d\mu_{\rm H}\,e^{-i\,
\frac{p\,q}{\hbar}}\,e^{i\,\frac{\mu\,p}{\hbar}} =\delta_{q,\mu}
\ea
That is, the basic objects in this representation are
Kronecker deltas. This is precisely what we had found in
Sec.~\ref{sec:2} for the B-type representation. How are now the
basic operators represented and what is the form of the inner
product? Regarding the operators, we expect that they are
represented in the opposite manner as in the previous
$p$-polarization case, but that they preserve the same features:
$\hat{p}$ does not exist (the derivative of the Kronecker delta is
ill defined), but its exponentiated version $\hat{V}(\lambda)$
does:
$$
\hat{V}(\lambda)\cdot\psi(q)=\psi(q+\lambda)
$$
and the operator $\hat{q}$ that now acts as multiplication has as its
eigenstates, the functions $\tilde{\psi}_\nu(q)=\delta_{\nu,q}$:
$$
\hat{q}\cdot\tilde{\psi}_\mu(q):=\mu\,\tilde{\psi}_\mu(q)
$$
What is now the nature of the quantum configurations space ${\cal Q}$? And
what is the measure thereon $\d\mu_q$? that defines the inner product we should
have:
$$
\langle\tilde{\psi}_\mu(q),\tilde{\psi}_\lambda(q)\rangle=\delta_{\mu,\lambda}
$$
The answer comes from one of the characterizations of the Bohr
compactification: we know that it is, in a precise sense, dual to the real
line but when equipped with the {\it discrete topology} $\R_\d$. Furthermore,
the measure on $\R_\d$ will be the `counting measure'. In this way we recover
the same properties we had for the previous characterization of the polymer
Hilbert space. We can thus write:
\be
\H_{{\rm poly},x}:=L^2(\R_\d,\d\mu_{\rm c})
\ee
This completes a precise construction of the B-type polymer representation
sketched in the previous section.
Note that if we had chosen  the opposite physical situation, namely that $q$,
the configuration observable, be the quantity that does not have a
corresponding operator, then we would have had the opposite realization: In
the $q$-polarization we would have had the type-A polymer representation and
the type-B for the $p$-polarization. As we shall see both scenarios have been
considered in the literature.

Up to now we have only focused our discussion on the kinematical aspects of
the quantization process. Let us now consider in the following section the
issue of dynamics and recall the approach that had been adopted in the
literature, before the issue of the removal of the regulator was reexamined
in \cite{CVZ}.

\section{Polymer Quantum Mechanics: Dynamics}
\label{sec:5}

As we have seen the construction of the polymer representation is
rather natural and leads to a quantum theory with different
properties than the usual Schr\"odinger counterpart such as its
non-separability, the non-existence of certain operators and the
existence of normalized eigen-vectors that yield a precise value
for one of the phase space coordinates. This has been done without
any regard for a Hamiltonian that endows the system with a
dynamics, energy and so on.

First let us consider the simplest case of a particle of mass $m$ in a
potential $V(q)$, in which the Hamiltonian $H$ takes the form,
\[
H=\frac{1}{2m} \,p^2+V(q)
\]
Suppose furthermore that the potential is given by a non-periodic
function, such as a polynomial or a rational function. We can
immediately see that a direct implementation of the Hamiltonian is
out of our reach, for the simple reason that, as we have seen, in
the polymer representation we can either represent $q$ or $p$, but
not both! What has been done so far in the literature? The
simplest thing possible: approximate the non-existing term by a
well defined function that {\it can} be quantized and hope for the
best. As we shall see in next sections, there is indeed more that
one can do.

At this point there is also an important decision to be made:
which variable $q$ or $p$ should be regarded as ``discrete"? Once
this choice is made, then it implies that the {\it other} variable
will not exist: if $q$ is regarded as discrete, then $p$ will not
exist and we need to approximate the kinetic term $p^2/2m$ by
something else; if $p$ is to be the discrete quantity, then $q$
will not be defined and then we need to approximate the potential
$V(q)$. What happens with a periodic potential? In this case one
would be modelling, for instance, a particle on a regular lattice
such as a phonon living on a crystal, and then the natural choice
is to have $q$ not well defined. Furthermore, the potential will
be well defined and there is no approximation needed.

In the literature both scenarios have been considered. For
instance, when considering a quantum mechanical system in
\cite{AFW}, the position was chosen to be discrete, so $p$ does
not exist, and one is then in the A type for the momentum
polarization (or the type B for the $q$-polarization). With this
choice, it is the kinetic term the one that has to be
approximated, so once one has done this, then it is immediate to
consider any potential that will thus be well defined. On the
other hand, when considering loop quantum cosmology (LQC), the
standard choice is that the configuration variable is not defined
\cite{lqc}. This choice is made given that LQC is regarded as the
symmetric sector of full loop quantum gravity where the connection
(that is regarded as the configuration variable) can not be
promoted to an operator and one can only define its exponentiated
version, namely, the holonomy. In that case, the canonically
conjugate variable, closely related to the volume,
becomes `discrete', just as in the full theory. This case is however,
different from the particle in a potential example. First we could
mention that the functional form of the
Hamiltonian constraint that implements dynamics has a
different structure, but the more important difference  lies in that
the system is constrained.

Let us return to the case of the particle in a potential and for
definiteness, let us start with the auxiliary kinematical framework in which:
$q$ is discrete, $p$ can not be
promoted and thus we have to approximate the kinetic term $\hat{p}^2/2m$.
How is this done? The standard prescription is to define, on the configuration
space ${\cal C}$, a regular `graph' $\gamma_{\mu_0}$. This consists of a
numerable set of points, equidistant, and characterized by a parameter $\mu_0$
that is the (constant) separation between points. The simplest example would
be to consider the set $\gamma_{\mu_0}=\{ q\in \R \; | \; q=n\,\mu_0\; ,
\forall \; n\in {\mathbb{Z}}\}$.

This means that the basic kets that will be considered
$|\mu_n\rangle$ will correspond precisely to labels $\mu_n$
belonging to the graph $\gamma_{\mu_0}$, that is,
$\mu_n=n\,\mu_0$. Thus, we shall only consider states of the form,
\be |\psi\rangle =\sum_n\,b_n\,|\mu_n \rangle\, .\label{graph-HS}
\ee This `small' Hilbert space $\H_{\gamma_{\mu_0}}$, the graph
Hilbert space, is a subspace of the `large' polymer Hilbert space
$\H_{\rm poly}$ but it is separable. The condition for a state of
the form (\ref{graph-HS}) to belong to the Hilbert space
$\H_{\gamma_{\mu_0}}$ is that the coefficients $b_n$ satisfy:
$\sum_n\,|b_n|^2 < \infty$.

Let us now consider the kinetic term $\hat{p}^2/2m$. We have to
approximate it by means of trigonometric functions, that can be
built out of the functions of the form $e^{i\lambda\,p/\hbar}$. As
we have seen in previous sections, these functions can indeed be
promoted to operators and act as translation operators on the kets
$|\mu\rangle$. If we want to remain in the graph $\gamma$, and not
create `new points', then one is constrained to considering
operators that displace the kets by just the right amount. That
is, we want the basic shift operator $\hat{V}(\lambda)$ to be such
that it maps the ket with label $|\mu_n\rangle$ to the next ket,
namely $|\mu_{n+1}\rangle$. This can indeed achieved by fixing,
once and for all, the value of the allowed parameter $\lambda$ to
be $\lambda=\mu_0$. We have then,
\[
\hat{V}(\mu_0)\cdot|\mu_n\rangle=|\mu_n+\mu_0\rangle=|\mu_{n+1}\rangle
\]
which is what we wanted. This basic `shift operator' will be the building
block for
approximating any (polynomial) function of $p$. In order to do that we notice
that
the function $p$ can be approximated by,
\[
p\approx
\frac{\hbar}{\mu_0}\,\sin\left(\frac{\mu_0\,p}{\hbar}\right)=\frac{\hbar}{2i\mu_0}
\left(e^{i\frac{\mu_0\,p}{\hbar}} -
e^{-i\frac{\mu_0\,p}{\hbar}}\right)
\]
where the approximation is good for $p<<\hbar/\mu_0$. Thus, one
can define a regulated operator $\hat{p}_{\mu_0}$ that depends on
the `scale' $\mu_0$ as:
\ba
\hat{p}_{\mu_0}\cdot|\mu_n\rangle &:=& \frac{\hbar}{2i\mu_0}[V(\mu_0)-V(-\mu_0)]
\cdot|\mu_n\rangle =\nonumber \\
&=& \frac{i\hbar}{2\mu_0}\,(|\mu_{n+1}\rangle -
|\mu_{n-1}\rangle) \ea
In order to regulate the operator
$\hat{p}^2$, there are (at least) two possibilities, namely to
compose the operator $\hat{p}_{\mu_0}$ with itself or to define a
new approximation. The operator $\hat{p}_{\mu_0}
\cdot\hat{p}_{\mu_0}$ has the feature that shifts the states two
steps in the graph to both sides. There is however another
operator that only involves shifting once:
\ba
\hat{p}^2_{\mu_0}\cdot|\nu_n\rangle &:=& \frac{\hbar^2}{\mu^2_0}\;
[2-\hat{V}(\mu_0)-
\hat{V}(-\mu_0)]\cdot|\nu_n\rangle=\nonumber \\
&=& \frac{\hbar^2}{\mu^2_0}\;(2|\nu_n\rangle-
|\nu_{n+1}\rangle-|\nu_{n-1}\rangle) \ea
which corresponds to the
approximation $p^2\approx
\frac{2\hbar^2}{\mu_0^2}(1-\cos(\mu_0\,p/\hbar))$, valid also in
the regime $p<<\hbar/\mu_0$. With these considerations, one can
define the operator $\hat{H}_{\mu_0}$, the Hamiltonian at scale
$\mu_0$, that in practice `lives' on the space
$\H_{\gamma_{\mu_0}}$ as,
\be \hat{H}_{\mu_0}:=\frac{1}{2m}\;\hat{p}^2_{\mu_0}+\hat{V}(q)\,
, \ee
that is a well defined, symmetric operator on
$\H_{\gamma_{\mu_0}}$. Notice that the operator is also defined on
$\H_{\rm poly}$, but there its physical interpretation is
problematic. For example, it turns out that the expectation value
of the kinetic term calculated on most states (states which are
not tailored to the exact value of the parameter $\mu_0$) is zero.
Even if one takes a state that gives ``reasonable`` expectation
values of the $\mu_0$-kinetic term and uses it to calculate the
expectation value of the kinetic term corresponding to a slight
perturbation of the parameter $\mu_0$ one would get zero. This
problem, and others that arise when working on  $\H_{\rm poly}$,
forces one to assign a physical interpretation to the Hamiltonian
$\hat{H}_{\mu_0}$ only when its action is restricted to the
subspace $\H_{\gamma_{\mu_0}}$.

Let us now explore the form that the Hamiltonian takes in the two
possible polarizations. In the $q$-polarization, the basis,
labelled by $n$ is given by the functions
$\chi_n(q)=\delta_{q,\mu_n}$. That is, the wave functions will
only have support on the set $\gamma_{\mu_0}$. Alternatively, one
can think of a state as completely characterized by the `Fourier
coefficients' $a_n$: $\psi(q) \leftrightarrow a_n$, which is the
value that the wave function $\psi(q)$ takes at the point
$q=\mu_n=n\,\mu_0$. Thus, the Hamiltonian takes the form of a
difference equation when acting on a general state $\psi(q)$.
Solving the time independent Schr\"odinger equation
$\hat{H}\cdot\psi=E\,\psi$ amounts to solving the difference
equation for the coefficients $a_n$.

The momentum polarization has a different structure. In this case, the
operator $\hat{p}^2_{\mu_0}$ acts as a multiplication operator,
\be
\hat{p}^2_{\mu_0}\cdot\psi(p)=\frac{2\hbar^2}{\mu_0^2}\left[1-\cos
\left(\frac{\mu_0\,p}{\hbar}\right)
\right]\,\psi(p)
\ee
The operator corresponding to $q$ will be represented as a derivative operator
$$
\hat{q}\cdot\psi(p):= i\hbar\;\partial_p\,\psi(p).
$$
For a generic potential $V(q)$, it has to be defined by means of
spectral theory defined now on a circle. Why on a circle? For
the simple reason that by restricting ourselves to a regular graph
$\gamma_{\mu_0}$, the functions of $p$ that preserve it (when
acting as shift operators) are of the form $e^{(i\, m\,
\mu_0\,p/\hbar)}$ for $m$ integer. That is, what we have are
Fourier modes, labelled by $m$, of period $2\pi\,\hbar/\mu_0$ in
$p$. Can we pretend then that the phase space variable $p$ is now
compactified? The answer is in the affirmative. The inner product
on periodic functions $\psi_{\mu_0}(p)$ of $p$ coming from the
full Hilbert space $\H_{\rm poly}$ and given by
\[
\langle \phi(p)|\psi(p)\rangle_{\rm poly}=
\lim_{L\mapsto\infty}\frac{1}{2L}\int_{-L}^{L}\d p\;
\overline{\phi}(p)\,\psi(p)
\]
is precisely equivalent to the inner product on the circle given by the
uniform measure
\[
\langle \phi(p)|\psi(p)\rangle_{\mu_0}=
\frac{\mu_0}{2\pi\hbar}\int_{-\pi\hbar/\mu_0}^{\pi\hbar/\mu_0}\d p\;
\overline{\phi}(p)\,\psi(p)
\]
with $p\in (-\pi\hbar/\mu_0,\pi\hbar/\mu_0)$. As long as one restricts
attention to the graph $\gamma_{\mu_0}$, one can work in this separable
Hilbert space $\H_{\gamma_{\mu_0}}$ of square integrable functions on $S^1$.
Immediately, one can see the limitations of this description. If the
mechanical system to be quantized is such that its orbits have values of
the momenta $p$ that are not small compared with $\pi\hbar/\mu_0$ then the
approximation taken will be very poor, and we don't expect neither the
effective classical description nor its quantization to be close to the
standard one. If, on the other hand, one is always within the region in
which the approximation can be regarded as reliable, then both classical
and quantum descriptions should approximate the standard description.

What does `close to the standard description' exactly mean needs, of course,
some further clarification. In particular one is assuming the existence of
the usual Schr\"odinger representation in which the system has a behavior
that is also consistent with observations. If this is the case, the natural
question is: How can we approximate such description from the polymer picture?
Is there a fine enough graph $\gamma_{\mu_0}$ that will approximate the system
in such a way that all observations are indistinguishable? Or even better,
can we define a procedure, that involves a refinement of the graph
$\gamma_{\mu_0}$ such that one recovers the standard picture?

It could also happen that a continuum limit can be defined but does not
coincide
with the `expected one'.
But there might be also physical systems for which there is {\it no}
standard description, or it just does not make sense.
Can in those cases the polymer representation,
if it exists, provide the correct physical description of the system under
consideration? For instance, if there exists a physical limitation to the
minimum scale set by $\mu_0$, as could be the case for a quantum theory of
gravity, then the polymer description would provide a true physical bound
on the value of certain quantities, such as $p$ in our example. This could
be the case for loop quantum cosmology, where there is a minimum
value for physical volume (coming from the full theory), and phase space
points near the `singularity' lie at the region where the approximation
induced by  the scale $\mu_0$ departs from the standard classical description.
If in that case the polymer quantum system is regarded as more
fundamental than the
classical system (or its standard Wheeler-De Witt quantization), then one
would interpret this discrepancies in the behavior as a signal of the
breakdown of classical description (or its `naive' quantization).

In the next section we present a method to remove the regulator $\mu_0$ which
was introduced as an intermediate step to construct the dynamics.
More precisely, we shall consider the construction of a continuum limit of
the polymer description by means of a renormalization procedure.

\section{The continuum limit}
\label{sec:6}

\noindent This section has two parts. In the first one we motivate
the need for a precise notion of the continuum limit of the
polymeric representation, explaining why the most direct, and
naive approach does not work. In the second part, we shall present
the main ideas and results of the paper \cite{CVZ}, where the
Hamiltonian and the physical Hilbert space in polymer quantum
mechanics are constructed as a continuum limit of effective
theories, following Wilson's renormalization group ideas. The
resulting physical Hilbert space turns out to be unitarily
isomorphic to the ordinary $\H_s=L^2(\R ,\d q)$ of the
Schr\"odinger theory.

Before describing the results of \cite{CVZ} we should discuss the
precise meaning of reaching a theory in the continuum. Let us for
concreteness consider the B-type representation in the
$q$-polarization. That is, states are functions of $q$ and the
orthonormal basis $\chi_\mu(q)$ is given by characteristic
functions with support on $q=\mu$. Let us now suppose we have a
Schr\"odinger state $\Psi(q)\in \H_s=L^2(\R ,\d q)$. What is the
relation between $\Psi(q)$ and a state in $\H_{{\rm poly},x}$? We
are also interested in the opposite question, that is, we would
like to know if there is a preferred state in $\H_s$ that is
approximated by an arbitrary state $\psi(q)$ in $\H_{{\rm
poly},x}$. The first obvious observation is that a Sch\"odinger
state $\Psi(q)$ does not belong to $\H_{{\rm poly},x}$ since it
would have an infinite norm. To see that note that even when the
would-be state can be formally expanded in the $\chi_\mu$ basis
as,
\[
\Psi(q) = \sum_{\mu}\;\Psi(\mu)\;\chi_{\mu}(q)
\]
where the sum is over the parameter $\mu \in \R$. Its associated
norm in  $\H_{{\rm poly},x}$ would be:
\[
|\Psi(q)|^2_{\rm poly}=\sum_{\mu} |\Psi(\mu)|^2 \rightarrow \infty
\]
which blows up. Note that in order to define a mapping
$P:\H_s\rightarrow \H_{{\rm poly},x}$, there is a huge ambiguity
since the  values of the function $\Psi(q)$ are needed in order to
expand the polymer wave function. Thus we can only define a
mapping in a dense subset ${\cal D}$ of $\H_s$ where the values of
the functions are well defined (recall that in $\H_s$ the value of
functions at a given point has no meaning since states are
equivalence classes of functions). We could for instance ask that
the mapping be defined for representatives of the equivalence
classes in $\H_s$ that are piecewise continuous. From now on, when
we refer to an element of the space $\H_s$ we shall be refereeing
to one of those representatives. Notice then that an element of
$\H_s$ {\it does} define an element of ${\rm Cyl}^*_\gamma$, the
dual to the space ${\rm Cyl}_\gamma$,  that is, the space of
cylinder functions with support on the (finite) lattice
$\gamma=\{\mu_1,\mu_2,\ldots,\mu_N\}$, in the following way:

\[
\Psi(q):{\rm Cyl}_\gamma \longrightarrow \C
\]
such that
\ba
\Psi(q)[\psi(q)]&=&(\Psi|\psi\rangle:= \sum_\mu
\overline{\Psi(\mu)}\;\langle\, \chi_{\mu} |\, \sum_{i=1}^N
\;\psi_i\,\chi_{\mu_i}\rangle_{{\rm poly}_\gamma}
\nonumber\\
& = &\sum_{i=1}^N
\;\overline{\Psi(\mu_i)}\,\psi_i \;<\;\infty
\ea
Note that this mapping could be seen as consisting of two parts:
First, a projection $P_\gamma:{\rm Cyl}^*\rightarrow {\rm
Cyl}_\gamma$ such that $P_\gamma(\Psi)=\Psi_\gamma(q):=
\sum_i\;\Psi(\mu_i)\,\chi_{\mu_i}(q)\in {\rm Cyl}_\gamma$. The
state $\Psi_\gamma$ is sometimes refereed to as the `shadow of
$\Psi(q)$ on the lattice $\gamma$'. The second step is then to
take the inner product between the shadow $\Psi_\gamma(q)$ and the
state $\psi(q)$ with respect to the polymer inner product $\langle
\Psi_\gamma|\psi\rangle_{{\rm poly}_\gamma}$. Now this inner
product is well defined. Notice that for any given lattice
$\gamma$ the corresponding projector $P_\gamma$
can be intuitively interpreted as some kind of `coarse graining
map' from the continuum to the lattice $\gamma$. In terms of
functions of $q$ the projection is replacing a continuous function
defined on $\R$ with a function over the lattice $\gamma\subset
\R$ which is a discrete set simply by restricting $\Psi$ to
$\gamma$. The finer the lattice the more points that we have on
the curve. As we shall see in the second part of this section,
there is indeed a precise notion of coarse graining that
implements this intuitive idea in a concrete fashion. In
particular, we shall need to replace the lattice $\gamma$ with a
decomposition of the real line in intervals (having the lattice
points as end points).

Let us now consider a system in the polymer representation in
which a particular lattice $\gamma_0$ was chosen, say with points
of the form $\{ q_k \in \R\,| q_k=ka_0\; ,\forall\, k \in \Z\}$,
namely a uniform lattice with spacing equal to $a_0$. In this
case, any Schr\"odinger wave function (of the type that we
consider) will have a unique shadow on the lattice $\gamma_0$. If
we refine the lattice $\gamma\mapsto \gamma_n$ by dividing each
interval in $2^n$ new intervals of length $a_n=a_0/2^n$ we have
new shadows that have more and more points on the curve.
Intuitively, by refining infinitely the graph we would recover the
original function $\Psi(q)$. Even when at each finite step the
corresponding shadow has a finite norm in the polymer Hilbert
space, the norm grows unboundedly and the limit can not be taken,
precisely because we can not embed $\H_s$ into $\H_{\rm poly}$.
Suppose now that we are interested in the reverse process, namely
starting from a polymer theory on a lattice and asking for the
`continuum wave function' that is best approximated by a wave
function over a graph. Suppose furthermore that we want to
consider the limit of the graph becoming finer. In order to give
precise answers to these (and other) questions we need to
introduce some new technology that will allow us to overcome these
apparent difficulties. In the remaining of this section we shall
recall these constructions for the benefit of the reader. Details
can be found in \cite{CVZ} (which is an application of the general
formalism discussed in \cite{MOWZ}).

The starting point in this construction is the concept of a scale
$C$, which allows us to define the effective theories and the
concept of continuum limit. In our case a scale is a decomposition
of the real line in the union of closed-open intervals, that cover
the whole line and do not intersect. Intuitively, we are shifting
the emphasis from the lattice points to the intervals defined by
the same points with the objective of approximating continuous
functions defined on $\R$ with functions that are constant on the
intervals defined by the lattice. To be precise, we define an
embedding, for each scale $C_n$ from ${\H}_{\rm poly}$ to $\H_s$
by means of a step function:
$$\sum_m\;\Psi(m a_n)\;\chi_{m a_n}(q)\quad\to\quad
\sum_m\;\Psi(m a_n)\;\chi_{\alpha_m}(q)\; \in \;\H_s
$$
with $\chi_{\alpha_n}(q)$ a characteristic function on the
interval $\alpha_m=[m a_n,(m+1)a_n)$. Thus, the shadows (living on
the lattice) were just an intermediate step in the construction of
the approximating function; this function is piece-wise constant
and can be written as a linear combination of step functions with
the coefficients provided by the shadows.

The challenge now is to define in an appropriate sense how one can
approximate all the aspects of the theory by means of this
constant by pieces functions. Then the strategy is that, for any
given scale, one can define an effective theory by approximating
the kinetic operator by a combination of the translation operators
that shift between the vertices of the given decomposition, in
other words by a periodic function in $p$. As a result one has a
set of effective theories at given scales which are mutually
related by coarse graining maps. This framework was developed in
\cite{CVZ}. For the convenience of the reader we briefly recall
part of that framework.

Let us denote the kinematic  polymer Hilbert space at the scale
$C_n$ as $\H_{C_n}$, and its basis elements as $e_{\a_i,C_n}$,
where $\a_i=[ia_n,(i+1)a_n)\in C_n$.
%
By construction this basis is orthonormal. The basis elements in
the dual Hilbert space $\H_{C_n}^*$ are denoted by
$\omega_{\a_i,C_n}$; they are also orthonormal. The states
$\omega_{\a_i,C_n}$ have a simple action on ${\rm Cyl}$,
$\omega_{\a_i,C_n} (\delta_{x_0,q})= \chi_{\a_i,C_n}(x_0)$. That
is, if $x_0$ is in the interval $\a_i$ of $C_n$ the result is one
and it is zero if it is not there.

Given any $m \leq n$, we define
$d^*_{m,n}:\H^*_{C_n}\to\H^*_{C_m}$ as the `coarse graining' map
between the dual Hilbert spaces, that sends the part of the
elements of the dual basis to zero while keeping the information
of the rest: $d^*_{m,n}(\omega_{\a_i,C_n})=\omega_{\beta_j,C_m}$
if $i=j2^{n-m}$, in the opposite case
$d^*_{m,n}(\omega_{\a_i,C_n})=0$.

At every scale the corresponding effective theory is given by the
hamiltonian $H_n$. These Hamiltonians will be treated as
quadratic forms, $h_n: \H_{C_n}\to\R$, given by \be h_n(\psi
)=\lambda_{C_n}^2(\psi ,H_n\psi )\, , \ee
 where
$\lambda_{C_n}^2$ is a normalizaton factor. We will see later that
this rescaling of the inner product is necessary in order to
guarantee the convergence of the renormalized theory. The
completely renormalized theory at this scale is obtained as \be
h_m^{\rm ren} := \lim_{C_n \to \R} d_{m,n}^\star h_n
.\label{cond-0} \ee
and the renormalized Hamiltonians are compatible with each other,
in the sense that
\[
d_{m,n}^\star h_n^{\rm ren} = h_m^{\rm ren} .
\]
In order to analyze the conditions for the convergence in
(\ref{cond-0}) let us express the Hamiltonian in terms of its
eigen-covectors end eigenvalues. We will work with effective
Hamiltonians that have a purely discrete spectrum (labelled by
$\nu$) $H_n\cdot\,\Psi_{\nu, C_n}=E_{\nu, C_n}\,\Psi_{\nu, C_n}$.
We shall also introduce, as an intermediate step, a cut-off in the
energy levels. The origin of this cut-off is in the approximation
of the Hamiltonian of our system at a given scale with a
Hamiltonian of a periodic system in a regime of small energies, as
we explained earlier. Thus, we can write \be h_m^{\nu_{\rm
cut-off}}  = \sum_{\nu =0}^{\nu_{\rm cut-off}} E_{\nu, C_m}
\Psi_{\nu, C_m} \otimes \Psi_{\nu, C_m} \, ,\label{diag} \ee where
the eigen covectors $\Psi_{\nu, C_m}$
are normalized according to the inner product rescaled by
$\frac{1}{\lambda_{C_n}^2}$, and the cut-off can vary up to a
scale dependent bound, $\nu_{\rm cut-off} \leq \nu_{\rm
max}(C_m)$. The Hilbert space of covectors together with such
inner product will be called $\H^{\star\rm{ren}}_{C_m}$.


In  the presence of a cut-off, the convergence of the
microscopically corrected Hamiltonians, equation (\ref{cond-0}) is
equivalent to the existence of the following two limits. The first
one is the convergence of the energy levels,
\be \lim_{C_n\rightarrow \R} E_{\nu,C_n} = E_{\nu}^{\rm ren}\,
.\label{cond-1} \ee
Second is the existence of the completely renormalized eigen
covectors,

\be \lim_{C_n\rightarrow \R}
d_{m,n}^\star\,\Psi_{\nu,C_n}=\Psi_{\nu,C_m}^{\rm
ren}\in\H^{\star\rm{ren}}_{C_m}\subset {\rm Cyl}^\star\,
.\label{cond-2} \ee
We clarify that the existence of the above limit means that
$\Psi_{\nu,C_m}^{\rm ren} (\delta_{x_0,q})$ is well defined for
any $\delta_{x_0,q} \in {\rm Cyl}$. Notice that this point-wise
convergence, if it can take place at all, will require the tuning
of the normalization factors $ \lambda_{C_n}^2$.

Now we turn to the question of the continuum limit of the
renormalized covectors. First we can ask for the existence of the
limit \be \label{PwiseCovector} \lim_{C_n\rightarrow \R}
\Psi_{\nu,C_n}^{\rm ren} (\delta_{x_0,q}) \ee for any
$\delta_{x_0,q} \in {\rm Cyl}$. When this limits exists there is a
natural action of the eigen covectors in the continuum limit.
Below we consider another notion of the continuum limit of the
renormalized eigen covectors.

When the completely renormalized eigen covectors exist, they form
a collection that is $d^\star$-compatible, $d_{m,n}^\star
\Psi_{\nu,C_n}^{\rm ren} = \Psi_{\nu,C_m}^{\rm ren}$. A sequence
of $d^\star$-compatible normalizable covectors define an element
of $\stackrel{\longleftarrow}{\H}_{\R}^{\star{\rm ren}}$, which is
the projective limit of the renormalized spaces of covectors \be
\stackrel{\longleftarrow}{\H}_{\R}^{\star{\rm ren}}:=
\stackrel{\longleftarrow}{\lim_{C_n\to \R}} \H_{C_n}^{\star{\rm
ren}} . \ee
The inner product in this space is defined by
\[
( \{ \Psi_{C_n} \} , \{ \Phi_{C_n}\} )^{\rm ren}_{\R} := \lim_{C_n
\to \R} ( \Psi_{C_n} ,  \Phi_{C_n} )^{\rm ren}_{C_n} .
\]
The natural inclusion of ${\cal C}^\infty_0$ in
$\stackrel{\longleftarrow}{\H}_{\R}^{\star{\rm ren}}$ {\em is by
an antilinear map} which assigns to any $\Psi \in {\cal
C}^\infty_0$ the $d^\star$-compatible collection $\Psi^{\rm
shad}_{C_n} := \sum_{\alpha_i} \omega_{\alpha_i} \bar{\Psi}
(L(\alpha_i)) \in \H_{C_n}^{\star{\rm ren}}\subset {\rm
Cyl}^\star$; $\Psi^{\rm shad}_{C_n}$ will be called the shadow of
$\Psi$ at scale $C_n$ and acts in ${\rm Cyl}$ as a piecewise
constant function. Clearly other types of test functions like
Schwartz functions are also naturally included in
$\stackrel{\longleftarrow}{\H}_{\R}^{\star{\rm ren}}$. In this
context a shadow is a state of the effective theory that
approximates a state in the continuum theory.

Since the inner product in
$\stackrel{\longleftarrow}{\H}_{\R}^{\star{\rm ren}}$ is
degenerate, the physical Hilbert space is defined as
\[
\H_{\rm phys}^\star
:=\stackrel{\longleftarrow}{\H}_{\R}^{\star{\rm ren}} / \ker
(\cdot ,\cdot )^{\rm ren}_{\R}
\]
\[
\H_{\rm phys} := \H_{\rm phys}^{\star \star}
\]

The nature of the physical Hilbert space, whether it is isomorphic
to the Schr\"odinger Hilber space, $\H_s$, or not, is determined
by the normalization factors $\lambda_{C_n}^2$ which can be
obtained from the conditions asking for compatibility of the
dynamics of the effective theories at different scales. The
dynamics of the system under consideration selects the continuum
limit.

Let us now return to the definition of the Hamiltonian in the
continuum limit. First consider the continuum limit of the
Hamiltonian (with cut-off) in the sense of its point-wise
convergence as a quadratic form. It turns out that if the limit of
equation (\ref{PwiseCovector}) exists for all the eigencovectors
allowed by the cut-off, we have $h_{\R}^{\nu_{\rm cut-off}{\rm
ren}} : \H_{{\rm poly},x} \to \R$ defined by \be h_{\R}^{\nu_{\rm
cut-off}{\rm ren}} (\delta_{x_0,q}) :=\lim_{C_n \to \R}
h_n^{\nu_{\rm cut-off}{\rm ren}} ([\delta_{x_0,q}]_{C_n}) .
\label{convcylstar} \ee
This Hamiltonian quadratic form in the continuum can be coarse
grained to any scale and, as can be expected, it yields the
completely renormalized Hamiltonian quadratic forms at that scale.
However, this is not a completely satisfactory continuum limit
because we can not remove the auxiliary cut-off ${\nu_{\rm
cut-off}}$. If we tried, as we include more and more
eigencovectors in the Hamiltonian the calculations done at a given
scale would diverge and doing them in the continuum is just as
divergent. Below we explore a more successful path.

We can use the renormalized inner product to induce an action of
the cut--off Hamiltonians on
$\stackrel{\longleftarrow}{\H}_{\R}^{\star{\rm ren}}$
\[
h_{\R}^{\nu_{\rm cut-off}{\rm ren}} (\{ \Psi_{C_n} \}) :=
\lim_{C_n \to \R} h_n^{\nu_{\rm cut-off}{\rm ren}} ((  \Psi_{C_n}
, \cdot )^{\rm ren}_{C_n}) ,
\]
where we have used the fact that $( \Psi_{C_n}, \cdot )^{\rm
ren}_{C_n} \in \H_{C_n}$. The existence of this limit is trivial
because the renormalized Hamiltonians are finite sums and the
limit exists term by term.

These cut-off Hamiltonians descend to the physical Hilbert space
\[
h_{\R}^{\nu_{\rm cut-off}{\rm ren}} ([\{ \Psi_{C_n} \}]) :=
h_{\R}^{\nu_{\rm cut-off}{\rm ren}} (\{ \Psi_{C_n} \})
\]
for any representative $\{ \Psi_{C_n} \} \in  [\{ \Psi_{C_n} \}]
\in \H_{\rm phys}^\star$.

Finally we can address the issue of removal of the cut-off. The
Hamiltonian $h_{\R}^{\rm ren} :
\stackrel{\longleftarrow}{\H}_{\R}^{\star{\rm ren}} \to \R$ is
defined by the limit
\[
h_{\R}^{\rm ren} := \lim_{\nu_{\rm cut-off} \to \infty}
h_{\R}^{\nu_{\rm cut-off}{\rm ren}}
\]
when the limit exists. Its corresponding Hermitian form in
$\H_{\rm phys}$ is defined whenever the above limit exists. This
concludes our presentation of the main results of \cite{CVZ}. Let
us now consider several examples of systems for which the
continuum limit can be investigated.

\section{Examples}
\label{sec:7}

In this section we shall develop several examples of systems that
have been treated with the polymer quantization. These examples
are simple quantum mechanical systems, such as the simple harmonic
oscillator and the free particle, as well as a quantum
cosmological model known as loop quantum cosmology.

\subsection{The Simple Harmonic Oscillator}

\noindent In this part, let us consider the example of a Simple
Harmonic Oscillator (SHO) with  parameters $m$ and $\omega$,
classically described by the following Hamiltonian
$$H=\frac{1}{2m}\,p^2+\frac{1}{2}\,m\,\omega^2\, x^2 .$$
Recall that from these parameters one can define a length scale
$D=\sqrt{\hbar/m\omega}$. In the standard treatment one uses this
scale to define a complex structure $J_D$ (and an inner product from
it), as we have described in detail that uniquely selects the standard
Schr\"odinger representation.

At scale $C_n$ we have an effective Hamiltonian for the Simple Harmonic
Oscillator (SHO) given by
\be
H_{C_n}=\frac{\hbar^2}{m a_n^2}\left[1-\cos{\frac{a_np}{\hbar}}
\right]+\frac{1}{2}\,m\,\omega^2 x^2\, .\label{sho}
\ee
If we interchange position and momentum, this Hamiltonian is exactly that
of a pendulum of mass $m$, length $l$ and subject to a constant
gravitational field $g$:
$$
\hat{H}_{C_n}=-\frac{\hbar^2}{2ml^2}\frac{d^2}{d\theta^2}+mgl(1-\cos{\theta})
$$
where those quantities are related to our system by,
$$l=\frac{\hbar}{m\,\omega\, a_n}\, ,\ \ \ g=\frac{\hbar\,\omega}{m\, a_n}\,
,\ \ \
\theta =\frac{p\, a_n}{\hbar}$$

That is, we are approximating, for each scale $C_n$ the SHO by a
pendulum. There is, however, an important difference. From our
knowledge of the pendulum system, we know that the quantum system
will have a spectrum for the energy that has two different
asymptotic behaviors, the SHO for low energies and the planar
rotor in the higher end, corresponding to oscillating and rotating
solutions respectively\footnote{Note that both types of solutions
are, in the phase space, closed. This is the reason behind the
purely discrete spectrum. The distinction we are making is between
those solutions inside the separatrix, that we call oscillating,
and those that are above it that we call rotating.}. As we refine
our scale and both the length of the pendulum and  the height of
the periodic potential increase, we expect to have an increasing
number of oscillating states (for a given pendulum system, there
is only a finite number of such states). Thus, it is justified to
consider the cut-off in the energy eigenvalues, as discussed in
the last section, given that we only expect a finite number of
states of the pendulum to approximate SHO eigenstates. With these
consideration in mind, the relevant question is whether the
conditions for the continuum limit to exist are satisfied. This
question has been answered in the affirmative in \cite{CVZ}. What
was shown there was that the eigen-values and eigen functions of
the discrete systems, which represent a discrete and
non-degenerate set, approximate those of the continuum, namely, of
the standard harmonic oscillator when the inner product is renormalized
by a factor $\lambda^2_{C_n}=1/2^n$.
This convergence implies that
the continuum limit exists as we understand it. Let us now
consider the simplest possible system, a free particle, that has
nevertheless the particular feature that the spectrum of the
energy is continuous.

\subsection{Free Polymer Particle}

\noindent In the limit $\omega\to 0$, the Hamiltonian of the
Simple Harmonic oscillator (\ref{sho}) goes to the Hamiltonian of
a free particle and the corresponding time independent
Schr\"odinger equation, in the $p-$polarization, is given by
$$
\left[ \,\frac{\hbar^2 }{m a_n^2} (1-\cos{\frac{a_np}{\hbar}})-E_{C_n}\,
\right]
\tilde{\psi} (p)=0
$$
where we now have that $p \in S^1$, with $p\in
(-\frac{\pi\hbar}{a_n}, \frac{\pi\hbar}{a_n})$. Thus, we have \be
E_{C_n}=\frac{\hbar^2}{m a_ n^2}
\left(1-\cos{\frac{a_np}{\hbar}}\right)\le E_{C_n, max}\equiv
2\frac{\hbar^2}{m a_ n^2} .\label{energy_free}
\ee
At each scale the energy of the particle we can describe is
bounded from above and the bound depends on the scale. Note that
in this case the spectrum is continuous, which implies that the
ordinary eigenfunctions of the Hilbert are not normalizable. This
imposes an upper bound in the value that the energy of the
particle can have, in addition to the bound in the momentum due to
its ``compactification''.

Let us first look for eigen-solutions to the time independent Schr\"odinger
equation, that is, for energy eigenstates. In the case of the ordinary free
particle, these correspond to constant momentum plane waves of the form
$e^{\pm(\frac{i p x}{\hbar})}$ and such that the ordinary
dispersion relation $p^2/2m=E$ is satisfied. These plane waves
are not square integrable and do not belong to the ordinary Hilbert space of
the
Schr\"odinger theory but they are still useful for extracting information about
the
system. For the polymer free particle  we have,
$$
\tilde{\psi}_{C_n} (p)=c_1 \delta (p-P_{C_n})+ c_2 \delta (p+P_{C_n})
$$
where $P_{C_n}$ is a solution of the previous equation considering a fixed
value
of $E_{C_n}$.
That is,
$$
P_{C_n}=P(E_{C_n})=\frac{\hbar}{a_n}\arccos{\left(1-\frac{m
a_n^2}{\hbar^2}\,E_{C_n}\right)}
$$
The inverse Fourier transform yields, in the `$x$ representation',
\ba
\psi_{C_n} (x_j)&=& \frac{1}{\sqrt{2\pi}}\int_{-\pi\hbar/a_n}^{\pi\hbar/a_n}
{\tilde{\psi}(p)\,e^{\frac{i\,a_n}{\hbar}\,p\,j}\,\d p}=\nonumber\\
&=& \frac{\hbar\,
\sqrt{2\pi}}{a_n}
\left(c_1 e^{ix_j P_{C_n}/\hbar }+c_2 e^{-ix_j P_{C_n}/\hbar}\right).
\label{FT}
\ea
with $x_j=a_n\,j$ for $j\in\Z$. Note that the eigenfunctions are
still delta functions (in the $p$ representation) and thus not
(square) normalizable with respect to the polymer inner product,
that in the $p$ polarization is just given by the ordinary Haar
measure on $S^1$, and there is no quantization of the momentum
(its spectrum is still truly continuous).

Let us now consider the time dependent Schr\"odinger equation,
$$
i\hbar\,\partial_t \,\tilde{\Psi}(p,t)=\hat{H}\cdot \tilde{\Psi}(p,t).
$$
Which now takes the form,
$$
\frac{\partial}{\partial t}\tilde{\Psi}(p,t)=\frac{-i\,\hbar}{m\,a_n}
\left(1-\cos{(a_n\,p/\hbar)}\right)\,\tilde{\Psi}(p,t)
$$
that has as its solution,
$$
\tilde{\Psi}(p,t)
=e^{-\frac{i\,\hbar}{m\,a_n}\left(1-\cos{(a_n\,p/\hbar)}
\right)\,t}\;\tilde{\psi}(p)
=e^{(-iE_{C_n}/ \hbar)\,t}\;\tilde{\psi}(p)
$$
for any initial function $\tilde{\psi}(p)$,
where $E_{C_n}$ satisfy the dispersion relation (\ref{energy_free}).
The wave function $\Psi(x_j,t)$, the $x_j$-representation of the wave
function, can be obtained for any given time $t$ by Fourier transforming
with (\ref{FT}) the wave function $\tilde{\Psi}(p,t)$.

In order to check out the convergence of the microscopically corrected
Hamiltonians we should analyze the convergence of the energy levels and of
the proper covectors.
In the limit $n\to\infty$, $E_{C_n}\to E=p^2/2m$ so we can be certain that
the eigen-values for the energy converge (when fixing the value of $p$).
Let us write the proper covector as
$\Psi_{C_n} = (\psi_{C_n}, \cdot )_{C_n}^{\rm ren} \in
\H^{\star{\rm ren}}_{C_n}$. Then
we can bring microscopic corrections to scale $C_m$ and look for
convergence of such corrections
$$\Psi^{\rm ren}_{C_m} \doteq \lim_{n \to \infty} d_{m,n}^\star \Psi_{C_n}.$$

It is easy to see that given any basis vector $e_{\alpha_i} \in
\H_{C_m}$ the following limit
\[
\Psi^{\rm ren}_{C_m}(e_{\alpha_i,C_m})  = \lim_{C_n \to \infty} \Psi_{C_n}(d_{n,m}(e_{\alpha_i,C_m}))
\]
exists and is equal to
\[
\Psi^{\rm shad}_{C_m}(e_{\alpha_i,C_m}) = [d^\star \Psi^{\rm Schr}](e_{\alpha_i,C_m})=
\Psi^{\rm Schr} (i a_m)
\]
where $\Psi^{\rm shad}_{C_m} $ is calculated using the free
particle Hamiltonian in the Schr\"odinger representation. This
expression defines the completely renormalized proper covector at
the scale $C_m$.

\subsection{Polymer Quantum Cosmology}

In this section we shall present a version of quantum cosmology
that we call polymer quantum cosmology. The idea behind this name
is that the main input in the quantization of the corresponding
mini-superspace model is the use of a polymer representation as
here understood. Another important input is the choice of
fundamental variables to be used and the definition of the
Hamiltonian constraint. Different research groups have made
different choices. We shall take here a simple model that has
received much attention recently, namely an isotropic, homogeneous
FRW cosmology with $k=0$ and coupled to a massless scalar field
$\varphi$. As we shall see, a proper treatment of the continuum
limit of this system requires new tools under development that are
beyond the scope of this work. We will thus restrict ourselves to
the introduction of the system and the problems that need to be
solved.

The system to be quantized corresponds to the phase space of
cosmological spacetimes that are homogeneous and isotropic and for
which the homogeneous spatial slices have a flat intrinsic
geometry  ($k=0$ condition). The only matter content is a
mass-less scalar field $\varphi$. In this case the spacetime
geometry is given by metrics of the form:
\[
\d s^2=-\d t^2 + a^2(t)\;(\d x^2+\d y^2+\d z^2)
\]
where the function $a(t)$ carries all the information and degrees
of freedom of the gravity part. In terms of the coordinates
$(a,p_a,\varphi,p_\varphi)$ for the phase space $\Gamma$ of the
theory, all the dynamics is captured in the Hamiltonian constraint
\[
{\cal C}:=-\frac{3}{8}\;\frac{p_a^2}{|a|}+8\pi
G\,\frac{p_\varphi^2}{2|a|^3}\approx 0
\]
The first step is to define the constraint on the kinematical
Hilbert space to find physical states and then a physical inner
product to construct the physical Hilbert space. First note that
one can rewrite the equation as:
\[
\frac{3}{8}\;p_a^2\,a^2 = 8\pi G\,\frac{p_\varphi^2}{2}
\]
If, as is normally done, one chooses $\varphi$ to act as an
internal time, the right hand side would be promoted, in the
quantum theory, to a second derivative. The left hand side is,
furthermore, symmetric in $a$ and $p_a$. At this point we have the
freedom in choosing the variable that will be quantized and the
variable that will not be well defined in the polymer
representation. The standard choice is that $p_a$ is not well
defined and thus, $a$ and any geometrical quantity derived from
it, is quantized. Furthermore, we have the choice of polarization
on the wave function. In this respect the standard choice is to
select the $a$-polarization, in which $a$ acts as multiplication
and the approximation of $p_a$, namely
$\sin({\lambda\,p_a})/\lambda$ acts as a difference operator on
wave functions of $a$. For details of this particular choice see
\cite{HW}. Here we shall adopt the opposite polarization, that is, we
shall have wave functions $\Psi(p_a,\varphi)$.

Just as we did in the previous cases, in order to gain intuition
about the behavior of the polymer quantized theory, it is
convenient to look at the equivalent problem in the classical
theory, namely the classical system we would get be approximating
the non-well defined observable ($p_a$ in our present case) by a
well defined object (made of trigonometric functions). Let us for
simplicity choose to replace $p_a\mapsto
\sin({\lambda\,p_a})/\lambda$. With this choice we get an
effective classical Hamiltonian constraint that depends on
$\lambda$:
\[
{\cal
C}_\lambda:=-\frac{3}{8}\;\frac{\sin(\lambda\,p_a)^2}{\lambda^2|a|}+8\pi
G\,\frac{p_\varphi^2}{2|a|^3}\approx 0
\]
We can now compute effective equations of motion by means of the
equations: $\dot{F}:=\{F,{\cal C}_\lambda\}$, for any observable
$F\in C^\infty(\Gamma)$, and where we are using the effective
(first order) action:
\[
S_\lambda=\int \d \tau (p_a\,\dot{a}+p_\varphi\,\dot{\varphi}-N\,{\cal
C}_\lambda)
\]
with the choice $N=1$. The first thing to notice is that the
quantity $p_\varphi$ is a constant of the motion, given that the
variable $\varphi$ is cyclic. The second observation is that
$\dot{\varphi}=8\pi\,G\frac{p_\varphi}{|a|^3}$ has the same sign
as $p_\varphi$ and never vanishes. Thus $\varphi$ can be used as a
(n internal) time variable. The next observation is that the
equation for $\left(\frac{\dot{a}}{a}\right)^2$, namely the
effective Friedman equation, will have a zero for a non-zero value
of $a$ given by
$$a^*=\frac{32\pi G}{3}\lambda^2 p_\varphi^2.$$
This is the value at which there will be bounce if the trajectory
started with a large value of $a$ and was contracting. Note that
the `size' of the universe when the bounce occurs depends on both
the constant $p_\varphi$ (that dictates the matter density) and
the value of the lattice size $\lambda$. Here it is important to
stress that for any value of $p_\varphi$ (that uniquely fixes the
trajectory in the $(a,p_a)$ plane), there {\it will} be a bounce.
In the original description in terms of Einstein's equations
(without the approximation that depends on $\lambda$), there in no
such bounce. If $\dot{a}<0$ initially, it will remain negative and
the universe collapses, reaching the singularity in a finite
proper time. What happens within the effective description if we
refine the lattice and go from $\lambda$ to
$\lambda_n:=\lambda/2^n$? The only thing that changes, for the
same classical orbit labelled by $p_\varphi$, is that the bounce
occurs at a `later time' and for a smaller value of $a^*$ but the
qualitative picture remains the same.

This is the main difference with the systems considered before. In
those cases, one could have classical trajectories that remained,
for a given choice of parameter $\lambda$, within the region where
$\sin(\lambda p)/\lambda$ {\it is} a good approximation to $p$. Of
course there were also classical trajectories that were outside
this region but we could then refine the lattice and find a new
value $\lambda'$ for which the new classical trajectory is well
approximated. In the case of the polymer cosmology, this is never
the case: {\it Every} classical trajectory will pass from a region
where the approximation is good to a region where it is not; this
is precisely where the `quantum corrections' kick in and the
universes bounces.

Given that in the classical description, the `original' and the
`corrected' descriptions are so different we expect that, upon
quantization, the corresponding quantum theories, namely the
polymeric and the Wheeler-DeWitt will be related in a non-trivial
way (if at all).

In this case, with the choice of polarization and for
a particular factor ordering we have,
\[
\left[\left(\frac{1}{\lambda}\sin(\lambda
p_a)\frac{\partial}{\partial
p_a}\right)^2+\frac{32\pi}{3}\,\ell^2_p\;
\frac{\partial^2}{\partial \varphi^2}\right]
\cdot\Psi(p_a,\varphi)=0
\]
as the Polymer Wheeler-DeWitt equation.

In order to approach the problem of the continuum limit of this
quantum theory, we have to
realize that the task is now somewhat different than before. This
is so given that the system is now a constrained system with a
constraint operator rather than a regular non-singular system with
an ordinary Hamiltonian evolution. Fortunately for the system
under consideration, the fact that the variable $\varphi$ can be
regarded as an internal time allows us to interpret the quantum
constraint as a generalized Klein-Gordon equation of the form
\[
\frac{\partial^2}{\partial \varphi^2}\;\Psi  = \Theta_\lambda\cdot \Psi
\]
where the operator $\Theta_\lambda$ is `time independent'. This
allows us to split the space of solutions into `positive and
negative frequency', introduce a physical inner product on the
positive frequency solutions of this equation and a set of
physical observables in terms of which to describe the system.
That is, one reduces in practice the system to one very similar to
the Schr\"odinger case by taking the positive square root of the
previous equation: $\frac{\partial{}}{\partial \varphi}\;\Psi  =
\sqrt{\Theta}_\lambda\cdot \Psi$. The question we are interested
is whether the continuum limit of these  theories (labelled by
$\lambda$) exists and whether it corresponds to the Wheeler-DeWitt
theory. A complete treatment of this problem lies, unfortunately,
outside the scope of this work and will be reported elsewhere
\cite{NETA}.

\section{Discussion}
\label{sec:8}

Let us summarize our results. In the first part of the article we
showed that the polymer representation of the canonical
commutation relations can be obtained as the limiting case of the
ordinary Fock-Schr\"odinger representation in terms of the
algebraic state that defines the representation. These limiting
cases can also be interpreted in terms of the naturally defined
coherent states associated to each representation labelled by the
parameter $d$, when they become infinitely `squeezed'. The two
possible limits of squeezing lead to two different polymer
descriptions that can nevertheless be identified, as we have also
shown, with the two possible polarizations for an abstract polymer
representation. This resulting theory has, however, very different
behavior as the standard one: The Hilbert space is non-separable,
the representation is unitarily inequivalent to the Schr\"odinger
one, and natural operators such as $\hat{p}$ are no longer well
defined. This particular limiting construction of
the polymer theory can shed some light for more complicated
systems such as field theories and gravity.

In the regular treatments of dynamics within the polymer
representation, one needs to introduce some extra structure, such
as a lattice on configuration space, to construct a Hamiltonian
and implement the dynamics for the system via a regularization
procedure. How does this resulting theory compare to the original
continuum theory one had from the beginning? Can one hope to
remove the regulator in the polymer description? As they
stand there is no direct relation or mapping from the polymer to a
continuum theory (in case there is one defined). As we have shown,
one can indeed construct in a systematic fashion such relation by
means of some appropriate notions related to the definition of a
scale, closely related to the lattice one had to introduce in the
regularization. With this important shift in perspective, and an
appropriate renormalization of the polymer inner product at each
scale one can, subject to some consistency conditions, define a
procedure to remove the regulator, and arrive to a Hamiltonian and
a Hilbert space.

As we have seen, for some simple examples such as a free particle
and the harmonic oscillator one indeed recovers the Schr\"odinger
description back. For other systems, such as quantum cosmological
models, the answer is not as clear, since the structure of the
space of classical solutions is such that the `effective
description' introduced by the polymer regularization at different
scales is qualitatively different from the original dynamics. A
proper treatment of these class of systems is underway and will be
reported elsewhere \cite{NETA}.

Perhaps the most important lesson that we have learned here is
that there indeed exists a rich interplay between the polymer
description and the ordinary Schr\"odinger representation. The
full structure of such relation still needs to be unravelled. We
can only hope that a full understanding of these issues will shed
some light in the ultimate goal of treating the quantum dynamics
of background independent field systems such as general
relativity.

\section*{Acknowledgments}

\noindent We thank A. Ashtekar, G. Hossain, T. Pawlowski and P.
Singh for discussions. This work was in part supported by CONACyT
U47857-F and 40035-F grants, by NSF PHY04-56913, by the Eberly
Research Funds of Penn State, by the AMC-FUMEC exchange program
and by funds of the CIC-Universidad Michoacana de San Nicol\'as de
Hidalgo.

\end{document}